\documentclass[10pt,conference]{IEEEtran}

\usepackage{epsfig}
\usepackage{graphicx}
\usepackage{cite}
\usepackage{amsmath}
\usepackage{amssymb,color}

\newcommand{\comment}[1]  {}
\def\BE{\begin{equation}}
\def\EE{\end{equation}}
\def\BEA{\begin{eqnarray}}
\def\EEA{\end{eqnarray}}

\DeclareMathOperator*{\argmax}{arg\,max}

\newtheorem{lemma}{Lemma}

\newcommand{\G}{\mathcal{G}}

\newcommand{\R}{\mathbb{R}}

\newcommand\ie{{\textsl{i.e.\,}}}
\newcommand\etc{{\textsl{etc.\,}}}
\newcommand\eg{{\textsl{e.g.\,}}}
\newcommand\etal{{\textsl{et al.\,\,}}}

\newcommand\vb{{b}}

\newcommand\vh{{h}}

\newcommand\vn{{n}}

\newcommand\vx{{x}}
\newcommand\vy{{y}}
\newcommand\vz{{z}}
\newcommand\mA{{A}} 

\newcommand\mI{{I}}
\newcommand\mJ{{J}}

\newcommand\mS{{S}}

\begin{document}

\title{Fixing Convergence of Gaussian Belief Propagation}

\author{\authorblockN{Jason K. Johnson}
\authorblockA{Center for Nonlinear Studies/T-4\\
Los Alamos National Laboratory\\
Los Alamos, NM 87545\\
Email: jasonj@lanl.gov}
\and
\authorblockN{Danny Bickson}
\authorblockA{IBM Haifa Research Lab\\
Mount Carmel, Haifa 31905, Israel\\
Email: dannybi@il.ibm.com}
\and
\authorblockN{Danny Dolev}
\authorblockA{School of Computer Science and Engineering\\
Hebrew University of Jerusalem\\
Jerusalem 91904, Israel\\
Email: dolev@cs.huji.ac.il}
}

\maketitle

\begin{abstract}

  Gaussian belief propagation (GaBP) is an iterative
  message-passing algorithm for inference in Gaussian
  graphical models. It is known that when GaBP converges it
  converges to the correct MAP estimate of the Gaussian 
  random vector and simple sufficient conditions for its 
  convergence have been established.

  In this paper we develop a double-loop algorithm for forcing
  convergence of GaBP. Our method computes the correct
  MAP estimate even in cases where standard GaBP would not have converged.
  We further extend this construction to compute least-squares
  solutions of over-constrained linear systems.
  We believe that our construction has numerous applications,
  since the GaBP algorithm is linked to solution of linear systems
  of equations, which is a fundamental problem in computer science and
  engineering.  As a case study, we discuss the linear detection
  problem. We show that using our new construction, we are able to
  force convergence of Montanari's linear detection algorithm, in cases
  where it would originally fail. As a consequence, we are able to
  increase significantly the number of users that can transmit
  concurrently.

\end{abstract}


\section{Introduction}

The Gaussian belief propagation algorithm is an efficient distributed
message-passing algorithm for inference over a Gaussian graphical model. GaBP is
also linked to the canonical problem of solving systems of linear
equations~\cite{MinSum,ISIT1,LSKalman}, one of the fundamental problems in
computer science and engineering, which explains the large number of algorithm
variants and applications. For example, the GaBP algorithm is applied for signal
processing~\cite{Loelinger07, Vontobel03, LSKalman, BibDB:FactorGraph,
Allerton08-2}, multiuser detection~\cite{BibDB:MontanariEtAl,ISIT2}, linear
programming~\cite{Allerton08-1}, ranking in social networks~\cite{PPNA08},
support vector machines~\cite{ECCS08} \etc Furthermore, it was recently shown
that some existing algorithms are specific instances of the GaBP algorithm,
including Consensus propagation~\cite{CP}, local probability
propagation~\cite{LPP}, multiuser detection~\cite{BibDB:MontanariEtAl},
Quadratic Min-Sum algorithm~\cite{MinSum}, Turbo decoding with Gaussian
densities~\cite{Turbo} and others. Two general sufficient conditions for
convergence of GaBP in loopy graphs are known: diagonal-dominance
\cite{BibDB:Weiss01Correctness} and walk-summability
\cite{BibDB:mjw_walksum_jmlr06}.  See also numerous studies in specific
settings~\cite{LPP, Turbo, BibDB:Weiss01Correctness, CP, BibDB:MontanariEtAl,
MinSum, BibDB:mjw_walksum_jmlr06, BibDB:jmw_walksum_nips}.

In this work, we propose a novel construction that fixes the convergence of the
GaBP algorithm, for any Gaussian model with positive-definite information matrix
(inverse covariance matrix), even when the currently known sufficient
convergence conditions do not hold. We prove that our construction converges to
the correct solution. Furthermore, we consider how this method may be used to
solve for the least-squares solution of general linear systems. As a specific
application, we discuss Montanari's multiuser detection
algorithm~\cite{BibDB:MontanariEtAl}. By using our construction we are able to
show convergence in practical CDMA settings, where the original algorithm did
not converge, supporting a significantly higher number of users on each cell.

This paper is organized as follows. Section \ref{sec:problem_settings}
outlines the problem model.  Section \ref{sec:GaBP} gives a brief introduction
to the GaBP algorithm. Section \ref{sec:our_const} describes our
novel double-loop construction for positive definite matrices. Section
\ref{sec_new_const} extends the construction for computing least-squares
solution of general linear systems.  We provide experimental results of
deploying our construction in the linear detection context in Section
\ref{sec:exp_results}. We conclude in Section \ref{sec:conc}.

\section{Problem setting}
\label{sec:problem_settings}

We wish to compute the \emph{maximum a posteriori} (MAP) estimate
of a random vector $x$ with Gaussian distribution (after conditioning
on measurements):
\begin{equation}
p(x) \propto \exp\{ -\tfrac{1}{2} x^T J x + h^T x \}\,, \label{eq:sys_prob}
\end{equation}
where $J \succ 0$ is a symmetric positive definite matrix (the information
matrix) and $h$ is the potential vector.  This problem is equivalent to solving
$J x = h$ for $x$ given $(h,J)$ or to solve the convex quadratic
optimization problem:
\begin{equation} \label{eq:objective_func}
\mbox{minimize} \;\; f(x) \triangleq \tfrac{1}{2} x^T J x - h^T x.
\end{equation}
We may assume without loss of generality (by rescaling variables) that
$J$ is normalized to have unit-diagonal, that is, $J \triangleq I-R$ with $R$
having zeros along its diagonal.  The off-diagonal entries of $R$ then
correspond to \emph{partial correlation coefficients} \cite{Lauritzen}. Thus,
the fill pattern of $R$ (and $J$) reflects the Markov structure
of the Gaussian distribution.  That is, $p(x)$ is Markov with respect to
the graph with edges $\G = \{(i,j) | r_{i,j} \neq 0 \}\,.$

If the model $J = I-R$ is \emph{walk-summable}
\cite{BibDB:jmw_walksum_nips,BibDB:mjw_walksum_jmlr06},
such that the spectral radius of $|R|=(|r_{ij}|)$ is
less than one ($\rho(|R|)<1$), then the method of GaBP may be used to solve this problem.  We
note that the walk-summable condition implies $I-R$ is positive definite.  An
equivalent characterization of the walk-summable condition is that $I-|R|$ is
positive definite.

\section{Gaussian belief propagation}
\label{sec:GaBP}

The Gaussian belief propagation algorithm  is an efficient
distributed message-passing algorithm for inference over a
Gaussian graphical model. Given the Gaussian density function \eqref{eq:sys_prob} or objective function \eqref{eq:objective_func},
we are interested in computing the MAP assignment:
\BE x^*  = \arg\max_x p(x) = \arg\min_x f(x) \nonumber \EE
The density $p(x)$ specifies a graphical model with respect to the
graph $G$ of the inverse covariance matrix $J$, 
with edge potentials (`compatibility functions') $\psi_{ij}$ and self-potentials
(`evidence') $\psi_{i}$. These graph potentials provide
a pairwise factorization of the Gaussian
distribution $p(x) \propto
\prod_{i=1}^{n}\psi_{i}(x_{i})\prod_{\{i,j\} \in G}\psi_{ij}(x_{i},x_{j}),$
        with $ \psi_{ij}(x_{i},x_{j})\triangleq
\exp(-x_{i}J_{ij}x_{j}),$ and
        $ \psi_{i}(x_{i}) \triangleq
        \exp\big(-\tfrac{1}{2} J_{ii}x_{i}^{2} + h_i x_i \big).$
 Then, we would like to calculate the
marginal densities, which must also be Gaussian,
\BE p(x_{i}) \sim
\mathcal{N}(\mu_{i}=(J^{-1} h)_{i},K_i \triangleq (J^{-1})_{ii})\,,
\nonumber \EE where $\mu_{i}$ and $K_{i}$ are the
marginal mean and variance,
respectively. The GaBP update rules are summarized in
Table~\ref{tab_summary}. We write $\mathbb{N}(i)$ to denote 
the set of neighbors of node $i$ in $G$.

\begin{table}
\centerline{ \small
\begin{tabular}{|c|c|l|}
  \hline
  \textbf{\#} & \textbf{Stage} & \textbf{Operation}\\
  \hline
  1. & \emph{Initialize} & Set $\alpha_{ij}=0$ and $\beta_{ij}=0$, $\forall (i,j) \in \G$\\ \hline
  2. & \emph{Iterate} & For all $(i,j) \in \G$\\
  & & \; $\alpha_{i \backslash j} = J_{ii} + \sum_{{k} \in \mathbb{N}(i) \backslash j} \alpha_{ki}$\\
  & & \; $\beta_{i \backslash j} = h_i + \sum_{k \in \mathbb{N}(i) \backslash j} \beta_{ki}$\\
  & & \; $\alpha_{ij} = -J_{ij}^2 \alpha_{i \backslash j}^{-1}$ \\
  & & \; $\beta_{ij} = -J_{ij} \alpha_{i \backslash j}^{-1} \beta_{i \backslash j}$\\
  & & end \\ \hline
  3. & \emph{Check} & If $\alpha$'s and $\beta$'s have converged,\\
  & &  continue to \#4. Else, return to \#2.\\\hline
  4. & \emph{Infer} & $\hat{K}_{i}=(J_{ii} + \sum_{{k} \in \mathbb{N}(i)} \alpha_{ki})^{-1}$ \\
& &  $\hat{\mu}_{i}= \hat{K}_i (h_i + \sum_{k \in \mathbb{N}(i)} \beta_{ki})$.\\
  \hline
  5. & \emph{Output} & $x^*_{i}= \hat{\mu}_{i}, \forall i.$ \\\hline
\end{tabular}}
\vspace{0.5cm}
\caption{Computing $x^*  = \argmax_{x} \exp(-\tfrac{1}{2}x^TJx + h^Tx)$ via GaBP.} \label{tab_summary}
\end{table}
\normalsize

It is known that if GaBP converges, it results in the exact
MAP estimate $x^*$, although the variance estimates $\hat{K}_i$ 
computed by GaBP are only approximations to the correct
variances \cite{BibDB:Weiss01Correctness}. 
The walk-summable condition guarantees that
GaBP converges \cite{BibDB:mjw_walksum_jmlr06}, generalizing the 
stricter condition \cite{BibDB:Weiss01Correctness} that
$J$ is diagonally dominant (\ie, $|J_{ii}|>\sum_{j\neq i}|J_{ij}| ,
\forall i$). An upper bound on convergence speed is 
given in \cite{Allerton08-1}.

\section{Our construction}

\label{sec:our_const}

This current paper presents a method to solve non-walksummable models,
where $J = I-R$ is positive definite but $\rho(|R|) \ge 1$, using
GaBP.  There are two key ideas: (1) using diagonal loading to create a
perturbed model $J'=J+\Gamma$ which is walk-summable (such that the GaBP
may be used to solve $J'x=h$ for any $h$) and (2) using this
perturbed model $J'$ and convergent GaBP algorithm as a
\emph{preconditioner} in a simple iterative method to solve the
original non-walksummable model.

\subsection{Diagonal Loading}

We may always obtain a walk-summable model by \emph{diagonal
loading}.  This is useful as we can then solve a related system
of equations efficiently using Gaussian belief propagation.
For example, given a non-walk-summable model $J = I-R$ we
obtain a related walk-summable model $J_\gamma = J + \gamma I$ that is
walk-summable for large enough values of $\gamma$:

\begin{lemma} Let $J=I-R$ and $J' \triangleq
J+\gamma I = (1+\gamma)I-R$.  Let $\gamma > \gamma^*$
where
\begin{equation}
\gamma^* = \rho(|R|)-1\,.
\end{equation}
Then, $J'$ is walk-summable and GaBP based on $J'$ converges.
\end{lemma}

\emph{Proof.} We normalize $J' = (1+\gamma) I - R$ to obtain
$J'_{\mathrm{norm}} = I - R'$ with $R' = (1+\gamma)^{-1} R$, which is
walk-summable if and only if $\rho(|R'|) < 1$.  Using $\rho(|R'|) =
(1+\gamma)^{-1} \rho(|R|)$ we obtain the condition $(1+\gamma)^{-1}
\rho(|R|) < 1$, which is equivalent to $\gamma >
\rho(|R|)-1$. $\diamond$

It is also possible to achieve the same effect by adding a general
diagonal matrix $\Gamma$ to obtain a walk-summable model.  For
example, for all $\Gamma > \Gamma^*$ where $\gamma^*_{ii} =
J_{ii}-\sum_{j\neq i} |J_{ij}|$ it holds that $J+\Gamma$ is
diagonally-dominant and hence walk-summable (see
\cite{BibDB:mjw_walksum_jmlr06}).
More generally, we could allow $\Gamma$ to be any symmetric positive-definite
matrix satisfying the condition $I+\Gamma \succ |R|$.  However, only
the case of diagonal matrices is explored in this present paper.

\subsection{Iterative Correction Method}

Now we may use the diagonally-loaded model $J' = J+\Gamma$ to solve
$Jx = h$ for any value of $\Gamma \ge 0$.
The basic idea here is to use the diagonally-loaded matrix $J' =
J+\Gamma$ as a \emph{preconditioner} for solving the $Jx = h$ using
the iterative method:
\begin{equation} \label{eq:iterative_method}
\hat{x}^{(t+1)} = (J+\Gamma)^{-1} (h+\Gamma \hat{x}^{(t)})
\end{equation}
Note that the effect of adding positive $\Gamma$ is to reduce the
size of the scaling factor $(J+\Gamma)^{-1}$ but we compensate for 
this damping effect by adding a feedback term $\Gamma \hat{x}$ to
the input $h$. Each step of this iterative method may also be 
interpreted as solving the following convex quadratic 
optimization problem based on the objective $f(x)$ from (\ref{eq:objective_func}):
\begin{equation}
\hat{x}^{(t+1)} = \arg\min_x \left\{ f(x) + \tfrac{1}{2} (x-x^{(t)})^T \Gamma
(x-x^{(t)}) \right\}
\end{equation}
This is basically a regularized version of Newton's method to minimize
$f(x)$ where we regularize the step-size
at each iteration. Typically, this regularization is used to
ensure positive-definiteness of the Hessian matrix when Newton's
method is used to optimize a non-convex function.  We instead use it to ensure that
$J+\Gamma$ is walk-summable, so that the update step can be
computed via Gaussian belief propagation.  Intuitively, this will
always move us closer to the correct solution, but slowly if $\Gamma$ is
large. It is simple to demonstrate the following:

\begin{lemma} Let $J \succ 0$ and $\Gamma \succeq 0$. Then,
$\hat{x}^{(t)}$ defined by (\ref{eq:iterative_method}) converges
to $x^* = J^{-1} h$ for all initializations $\hat{x}^{(0)}$.
\end{lemma}

\emph{Comment.} The proof is given for a general (non-diagonal)
$\Gamma \succeq 0$.  For diagonal matrices, this
is equivalent to requiring $\Gamma_{ii} \ge 0$ for $i=1,\dots,n$.

\emph{Proof.} First, we note that there is only one possible
fixed-point of the algorithm and this is $x^* = J^{-1} h$.
Suppose $\bar{x}$ is a fixed point: $\bar{x} = (J+\Gamma)^{-1}
(h+\Gamma \bar{x})$.  Hence, $(J+\Gamma) \bar{x} = h+\Gamma \bar{x}$
and $J \bar{x} = h$. For non-singular $J$, we must then have $\bar{x}
= J^{-1} h$. Next, we show that the method converges.  Let $e^{(t)} =
\hat{x}^{(t)}-x^*$ denote the error of the $k$-th estimate.  The
error dynamics are then $e^{(t+1)} = (J+\Gamma)^{-1} \Gamma e^{(t)}$.
Thus, $e^{(t)} = ((J+\Gamma)^{-1}\Gamma)^k e^{(0)}$ and the error
converges to zero if and only if $\rho((J+\Gamma)^{-1} \Gamma) < 1$, or
equivalently $\rho(H)<1$ where $H = (J+\Gamma)^{-1/2} \Gamma
(J+\Gamma)^{-1/2} \succeq 0$ is a symmetric positive semi-definite
matrix.  Thus, the eigenvalues of $H$ are non-negative and we must
show that they are less than one.  It is simple to check that if
$\lambda$ is an eigenvalue of $H$ then $\frac{\lambda}{1-\lambda}$ is
an eigenvalue of $\Gamma^{1/2} J^{-1} \Gamma^{1/2} \succeq 0$.  This
is seen as follows: $Hx = \lambda x$, $(J+\Gamma)^{-1} \Gamma y =
\lambda y$ ($y = (J+\Gamma)^{-1/2} x$), $\Gamma y = \lambda
(J+\Gamma)y$, $(1-\lambda) \Gamma y = \lambda J y$, $J^{-1} \Gamma y =
\frac{\lambda}{1-\lambda} y$ and $\Gamma^{1/2} J^{-1} \Gamma^{1/2} z =
\frac{\lambda}{1-\lambda} z$ ($z = \Gamma^{1/2} y$) [note that $\lambda \neq
1$, otherwise $Jy=0$ contradicting $J \succ 0$].  Therefore
$\frac{\lambda}{1-\lambda} \ge 0$ and $0 \le \lambda < 1$.  Then
$\rho(H) < 1$, $e^{(t)} \rightarrow 0$ and $\hat{x}^{(t)} \rightarrow
  x^*$ completing the proof. $\diamond$

Now, provided we also require that $J' = J+\Gamma$ is walk-summable, we may
compute $x^{(t+1)} = (J+\Gamma)^{-1} h^{(t+1)}$, where $h^{(t+1)} = h +
\Gamma \hat{x}^{(t)}$, by performing Gaussian belief propagation to
solve $J' x^{(t+1)} = h^{(t+1)}$.  Thus, we obtain a double-loop
method to solve $Jx = h$.  The inner-loop performs GaBP and the outer-loop
computes the next $h^{(t)}$. The overall procedure converges provided
the number of iterations of GaBP in the inner-loop is made large enough
to ensure a good solution to $J' x^{(t+1)} = h^{(t+1)}$.
Alternatively, we may compress this double-loop procedure
into a single-loop procedure by preforming just one
iteration of GaBP  message-passing per iteration of the outer loop.
Then it may become necessary to use the following damped update of $h^{(t)}$
with step size parameter $s \in (0,1)$:
\begin{eqnarray}
h^{(t+1)} &=& (1-s) h^{(t)} + s (h + \Gamma \hat{x}^{(t)}) \nonumber\\
&=& h + \Gamma ((1-s) \hat{x}^{(t-1)} + s \hat{x}^{(t)})
\end{eqnarray}
This single-loop method converges for sufficiently small values
of $s$.  In practice, we have found good convergence with $s=\frac{1}{2}$.
This single-loop method can be more efficient than the double-loop
method.

\section{Extension to General Linear  Systems}

\label{sec_new_const} In this section, we efficiently extend the
applicability of the proposed double-loop construction for a general linear system of
equations (possibly over-constrained.)
Given a full column rank matrix $\tilde{J} \in \R^{n \times k}$, $n \ge k$,
and a shift vector $\tilde{h}$, we are interested in solving the
least squares problem $\min_x ||\tilde{J}x - \tilde{h}||^2_2$.  The naive
approach for using GaBP would be to take the information matrix
$\bar{J} \triangleq (\tilde{J}^T\tilde{J})$, and the shift vector
$\bar{h} \triangleq \tilde{J}^T\tilde{h}$.  Note that $\bar{J}$ is
positive definite and we can use GaBP to solve it.  The
MAP solution is \BE x =
\bar{J}^{-1}\bar{h} =(\tilde{J}^T\tilde{J})^{-1}\tilde{J}h\,,\label{pi} \EE which is the
pseudo-inverse solution.

Note, that the above construction has two drawbacks: first, we need
to explicitly compute $\bar{J}$ and $\bar{h}$, and second, $\bar{J}$
may not be sparse in case the original matrix $\tilde{J}$ is
sparse. To overcome this problem, following \cite{ISIT2}, we construct
a new symmetric data matrix $\bar{\bar{\mJ}}$ based on the arbitrary
rectangular matrix
$\tilde{\mJ}\in\mathbb{R}^{n\times k}$ \BE
\label{newJ} \bar{\bar{\mJ}}\triangleq\left(
  \begin{array}{cc}
    \mI_{k \times k} & \tilde{\mJ}^T \\
    \tilde{\mJ} & \mathbf{0}_{n \times n} \\
  \end{array}
\right)\in\mathbb{R}^{(k+n)\times(k+n)}\,.\nonumber \EE Additionally, we
define a new hidden variable vector
$\tilde{\vx}\triangleq\{x^{T},\vz^{T}\}^{T}\in\mathbb{R}^{(k+n)}$,
where $\vx\in\mathbb{R}^{k}$ is the solution vector and
$\vz\in\mathbb{R}^{n}$ is an auxiliary hidden vector, and a new
shift vector $\bar{\bar{\vh}}\triangleq\{\mathbf{0}_{k \times
  1}^{T},\vh^{T}\}^{T}\in\mathbb{R}^{(k+n)}$. 

\begin{lemma}
Solving $\bar{\bar{x}} = \bar{\bar{J}}^{-1}\bar{\bar{h}}$ and taking the first $k$ entries
is identical to solving Eq. \ref{pi}.
\end{lemma}
\emph{Proof.}
Is given in \cite{ISIT2}.

 For applying our double-loop construction on the new system $(\bar{\bar{h}},\bar{\bar{J}})$ to obtain the
solution to Eq.~(\ref{pi}), we need to confirm that the matrix $\bar{\bar{J}}$ is positive definite. (See lemma 2).
To this end, we add a diagonal weighting $-\gamma I$ to the lower right block:
 \BE
 \hat{\mJ}\triangleq\left(
  \begin{array}{cc}
    \mI_{k \times k} & \tilde{\mJ}^T \\
    \tilde{\mJ} & -\gamma I_{n \times n} \\
  \end{array}
\right)\in\mathbb{R}^{(k+n)\times(k+n)}\,.\nonumber \EE 
Then we rescale $\hat{J}$ to make it unit diagonal (to deal with the negative sign of the lower right block we use a complex Gaussian notation as done in \cite{BibDB:MontanariEtAl}). It is clear for a large enough $\gamma$ we are left with a walk-summable model, where the rescaled $\hat{J}$ is a hermitian positive definite matrix and $\rho(|\hat{J}-I|)<1$. Now it is possible to use the double-loop technique to compute Eq. \ref{pi}. Note that adding $-\gamma I$ to the lower right block of $\hat{J}$ is equivalent to adding $\gamma I$ into Eq. 7:
\BE x = (\tilde{J}^T\tilde{J} +  \gamma I)^{-1}\tilde{J}^Th\, \label{hatj} \EE
where $\gamma$ can be interpreted as a regularization parameter.

\section{Experimental results}
\label{sec:exp_results}

\subsection{Linear detection in linear channels}

Consider a discrete-time channel with a real input vector
\mbox{$\vx=\{x_{1},\ldots,x_{K}\}^{T}$} governed by an arbitrary
prior distribution, $P_{\vx}$, and a corresponding real output
vector
\mbox{$\vy=\{y_{1},\ldots,y_{K}\}^{T}=f\{\vx^{T}\}\in\mathbb{R}^{K}$}.
Here, the function $f\{\cdot\}$ denotes the channel
transformation. By definition, linear detection compels the
decision rule to be\BE\label{eq_gld}
\hat{\vx}=\Delta\{\vx^{\ast}\}=\Delta\{\mA^{-1}\vb\}\,, \EE where
$\vb=\vy$ is the $K\times 1$ observation vector and the $K\times
K$ matrix $\mA$ is a positive-definite symmetric matrix
approximating the channel transformation. The vector $\vx^{\ast}$
is the solution (over $\mathbb{R}$) to $\mA\vx=\vb$. Estimation is
completed by adjusting the (inverse) matrix-vector product to the
input alphabet, dictated by $P_{\vx}$, accomplished by using a
proper clipping function $\Delta\{\cdot\}$ (\eg, for binary
signaling $\Delta\{\cdot\}$ is the sign function).

For example, linear channels, which appear extensively in many
applications in communication and data storage systems, are
characterized by the linear relation\[ \vy=f\{\vx\}=C\vx+\vn\,,
\] where $\vn$ is a $K\times 1$ additive noise vector and
\mbox{${C}=\mS^{T}\mS$} is a positive-definite symmetric matrix,
often known as the correlation matrix. The $N\times K$ matrix
$\mS$ describes the physical channel medium while the vector $\vy$
corresponds to the output of a bank of filters matched to the
physical channel $\mS$.

Assuming linear channels with AWGN with variance $\sigma^{2}$ as the
ambient noise, the linear minimum mean-square error (MMSE) detector
can be described by using \mbox{$\mA=C+\sigma^{2}\mI_{K}$}, known
to be optimal when the input distribution $P_{\vx}$ is Gaussian.
In general, linear detection is suboptimal because of its
deterministic underlying mechanism (\ie, solving a given set of
linear equations), in contrast to other estimation schemes, such
as MAP or maximum likelihood, that emerge from an optimization
criteria.

\subsection{Montanari's iterative algorithm for computing the MMSE detector}

Recent work by Montanari \etal ~\cite{BibDB:MontanariEtAl} introduces
an efficient iterative algorithm for computing the MMSE
detector. Following this work, Bickson \etal showed that this
algorithm is an instance of the GaBP algorithm~\cite{ISIT2}.

In the current work, we apply our novel technique for forcing the
convergence of Montanari's algorithm. To remind, Montanari's algorithm computes the MMSE solution
\[ x = (C+\sigma^2I_K)^{-1}y\,. \]

We use the following setting:
given a random-spreading CDMA code with chip sequence length $n =
256$, and $k = 64$ users. We assume a diagonal AWGN with $\sigma^2 =
1$. Matlab code of our implementation is available on
\cite{MatlabGABP}.

Using the above settings, we have drawn at random random-spreading CDMA matrix.
Typically, the
sufficient convergence conditions for the GaBP algorithm do not
hold. For example, we have drawn at random a randomly-spread CDMA
matrix with $\rho(|I_{K}-{C^{N}}|) = 4.24$, where $C^N$ is a diagonally-normalized version of $(C+\sigma^2 I_K)$. Since $\rho(|I_K-{C^{N}}|) > 1$, the GaBP
algorithm for multiuser detection is not guaranteed to converge.

\begin{figure}[ht!]
\centering{
  \includegraphics[bb=83 180 504 488,scale=0.49,clip]{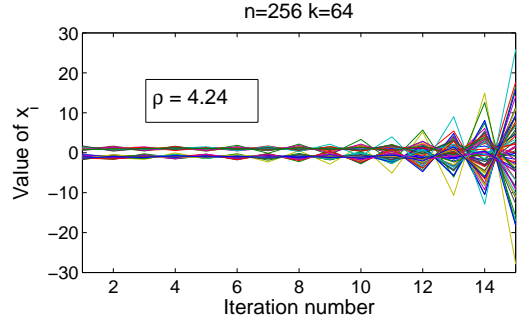}\\
  \caption{Divergence of the GaBP algorithm for the multiuser detection problem,
when $n=256, k=64$. }\label{fig:residual_gabp}
}
\end{figure}

Figure~\ref{fig:residual_gabp} shows that under the above settings,
the GaBP algorithm indeed diverged. The $x$-axis represent iteration
number, while the values of different $x_i$ are plotted using
different colors. This figure depicts well the fluctuating divergence
behavior.

\begin{figure}[h!]
\centering{
  \includegraphics[scale=0.44,clip]{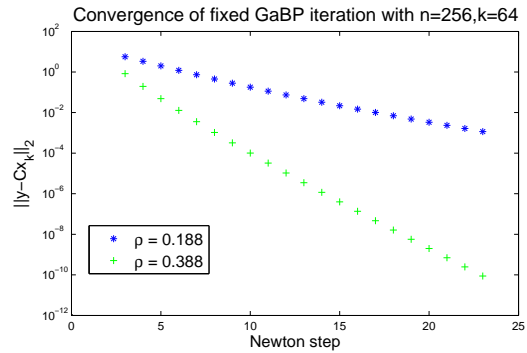}\\
  \caption{Convergence of the fixed GaBP iteration under the same settings
($n=256, k=64$)}\label{fig:newton}
}
\end{figure}

Next, we deployed our proposed construction and used a diagonal
loading to force convergence.  Figure~\ref{fig:newton} shows two
different possible diagonal loadings. The $x$-axis shows the Newton step
number, while the $y$-axis shows the residual. We experimented with two
options of diagonal loading.  In the first, we forced the matrix to be
diagonally-dominant (DD). In this case, the spectral radius $\rho =
0.188$. In the second case, the matrix was not DD, but the spectral
radius was $\rho = 0.388$. Clearly, the Newton method converges
faster when the spectral radius is larger. In both cases the inner iterations
converged
in five steps to an accuracy of $10^{-6}$.

The tradeoff between the amount of diagonal weighting to the total convergence speed is shown in Figures 3,4. A CDMA multiuser detection problem is shown ($k=128$, $n=256$). Convergence threshold for the inner and outer loops where $10^{-6}$ and $10^{-3}$. The $x$-axis present the amount of diagonal weighting normalized such that 1 is a diagonally-dominant matrix. $y$-axis represent the number of iterations. As expected, the outer loop number of iterations until convergence grows with the increase of $\gamma  $. In contrast, the average number of inner loop iterations per Newton step (Figure 4)\ tends to decrease as $\gamma$ increases. The total number of iterations (inner $\times$ outer) represents the tradeoff between the inner and outer iterations and has a clear global minima.\begin{figure}[h!]
\centering{
  \includegraphics[scale=0.44,clip]{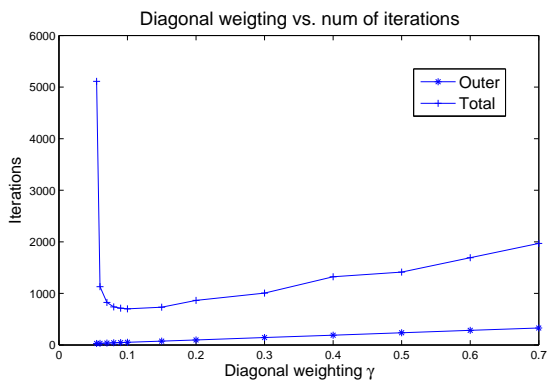}\\
  \caption{Effect of diagonal weighting on outer loop convergence speed. }\label{fig:outer}
}
\end{figure}

\begin{figure}[h!]
\centering{
  \includegraphics[scale=0.44,clip]{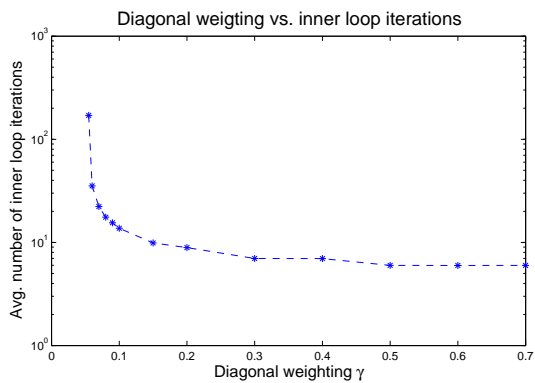}\\
  \caption{Effect of diagonal weighting on inner loop convergence speed.}\label{fig:inner}
}
\end{figure}

\section{Conclusions and Future Work}
\label{sec:conc}

We have presented an iterative method based on Gaussian belief
propagation which always converges to the correct global solution,
even in models where Gaussian belief propagation alone does
not converge. Essentially, this involves adding a diagonal-loading
term to force the model to become walk-summable such that GaBP
converges in this modified model and adding a feedback
mechanism that corrects the damping caused by the
diagonal-loading term.

We believe that there are numerous applications for our construction
in many domains, since GaBP is related to the solution of linear
systems of equations. As an example, we discuss the case of multiuser
detection.  We gave a concrete example, where a state-of-the-art
linear iterative algorithm for detection fails to converge. Using our
construction we are able to force convergence for computing the
correct MMSE detector.

There are a number of directions for further development.
Most importantly, it would be very useful to develop a simple method
to select $\Gamma$ so as to optimize the rate of convergence of the
overall method.  In the double-loop method, it is seen that there
is a trade-off in deciding how large $\Gamma$ should be. For larger
$\Gamma$ (beyond the threshold of walk-summability) GaBP converges
faster by accelerating the inner-loop of our algorithm.  However, larger
$\Gamma$ will also make the outer-loop converge more slowly.  Hence,
we must somehow balance these competing objectives in choosing $\Gamma$.
In the single-loop method, it would be useful to develop an adaptive
method to optimize the step-size parameter $s$.  Lastly, it may also
prove useful to exploit a more general class of perturbations
beyond the diagonal-loading method used in this paper.
\section*{Acknowledgment}
Danny Dolev is Incumbent of the Berthold Badler Chair in Computer Science. Danny Dolev was  supported in part by the Israeli Science Foundation (ISF) Grant number 0397373.

\bibliographystyle{IEEEtran}   
\bibliography{ISIT09-1}       

\end{document}